# Ferromagnetic order in U(Rh,Co)Ge

N.T. Huy and A. de Visser

*Van der Waals-Zeeman Institute, University of Amsterdam*
*Valckenierstraat 65, 1018 XE Amsterdam, The Netherlands*

Abstract:

We report the variation of ferromagnetic order in the pseudo-ternary compounds URh$_{1-x}$Co$_x$Ge ($0 \leq x \leq 1$). Magnetization and transport data taken on polycrystalline samples show that the Curie temperature $T_C$ gradually increases with increasing Co content from a value of 9.5 K for URhGe to a maximum value of 20 K for $x = 0.6$ and then steadily decreases to 3 K for UCoGe. The magnetic interaction strength varies smoothly across the series. For all samples the electrical resistivity for $T < T_C$ follows the behaviour $\rho = \rho_0 + AT^2$. The $A$ coefficient is dominated by scattering at spin waves and is strongly enhanced for $x = 0$ and 1.

Corresponding author:
  Dr. A. de Visser
  Van der Waal-Zeeman Institute, University of Amsterdam
  Valckenierstraat 65, 1018 XE Amsterdam, The Netherlands
  Phone: +31-(0)205255732
  Fax: +31-(0)205255788
  E-mail: a.devisser@uva.nl

# 1. Introduction

The intermetallic compounds URhGe [1] and UCoGe [2] belong to the family of ferromagnetic superconductors, which attracts much attention. Metallic ferromagnetism is observed upon cooling to liquid helium temperatures, and at even lower temperatures superconductivity sets in, which coexists with ferromagnetic order. This coexistence is highly unusual, because in the standard BCS theory superconductivity and ferromagnetism are competing ground states: the exchange field due to long range ferromagnetic order impedes the formation of phonon-mediated spin-singlet Cooper pairs [3]. Evidence is at hand, however, that in ferromagnetic superconductors an alternative pairing mechanism is at work: magnetic fluctuations associated with a magnetic quantum critical point mediate spin-triplet Cooper pairs [4]. The family of ferromagnetic superconductors is small: $UGe_2$ (under pressure) [5], URhGe [1], UIr (under pressure) [6] and UCoGe [2]. In these materials ferromagnetic order has a strong itinerant character. The exchange split Fermi surface consists of majority and minority spin sheets. The Cooper pair interaction may be attractive on both Fermi surface sheets and consequently a superconducting condensate with both equal-spin pairing Cooper states $|\uparrow\uparrow\rangle$ and $|\downarrow\downarrow\rangle$ may form. Thus, in principle ferromagnetic superconductors are two-band $p$-wave superconductors [4,7,8].

The compounds URhGe and UCoGe both crystallize in the orthorhombic TiNiSi structure (space group $P_{nma}$) [9,10]. The Curie temperatures are 9.5 K [11] and 3.0 K [2], respectively, and superconductivity is found below 0.25 K [1] and 0.8 K [2], respectively. Magnetism is stronger in URhGe. The ordered magnetic moment, $m_0$, amounts to 0.20 $\mu_B$/U-atom (powder averaged value) [1,11,12] and the magnetic entropy, $S_{mag}$, calculated by integrating the 5$f$ electron specific heat $\int C_{5f}/TdT$ up to a temperature of approximately $1.5 \times T_C$ amounts to $0.46 \times R\ln 2$ (here $R$ is the gas constant) [12,13]. This, together with the small ratio $m_0/p_{eff} \sim 0.1$ (the effective moment $p_{eff} \sim 1.7$ $\mu_B$/U-atom [11,12]) confirms the itinerant nature of the ferromagnetic state. For UCoGe the weak ordered moment $m_0 = 0.02$-$0.03$ $\mu_B$/U-atom and the much reduced value of the magnetic entropy, $S_{mag} = 0.04 \times R\ln 2$, reveal the proximity to a ferromagnetic quantum critical point [2].

Previously, we reported the evolution of ferromagnetic order in URhGe upon replacing Rh by Ru or Co, or Ge by Si [14]. This work was motivated by the possibility to attain a ferromagnetic quantum critical point in URhGe by doping. Indeed in the URh$_{1-x}$Ru$_x$Ge series $T_C(x) \to 0$ for $x = 0.38$ [13]. Notice, in the literature it was initially reported that UCoGe has a paramagnetic ground state [11,15]. Magnetization data taken on polycrystalline samples with $x$ = 0.0, 0.2, 0.4, 0.6, 0.8 and 0.9 revealed that the Curie temperature ($T_C$ = 9.5 K for $x$ = 0.0) increases up to $T_C$ = 20 K for $x$ = 0.6, and then drops to 8.0 K for $x$ = 0.9 [14]. These data hinted at a ferromagnetic instability for $x \lesssim 1.0$, but later experiments surprisingly showed ferromagnetism survives in the URh$_{1-x}$Co$_x$Ge alloys up to $x$ = 1.0 [2]. Since Rh and Co atoms are isoelectronic, the variation of $T_C$ as a function of $x$, with the broad maximum near $x$ = 0.6, can be qualitatively understood on the basis of a Doniach-like diagram [16]. However, the anisotropic variation of the lattice parameters results in strong anisotropies in the hybridization phenomena, which hampers a quantitative analysis [14].

In this work we report magnetization measurements conducted to investigate the depression of the ferromagnetic state for $x > 0.9$, as well as transport measurements across the whole URh$_{1-x}$Co$_x$Ge series. We find that the itinerant nature of the ferromagnetic state is preserved over the whole concentration range. The magnetic interaction strength varies smoothly across the series. Transport measurements show that the coefficient $A$ of the $T^2$ term in the electrical resistivity is strongly enhanced towards $x$ = 0.0 and 1.0.

## 2. Sample preparation and experiments

The polycrystalline URh$_{1-x}$Co$_x$Ge samples were prepared by arc-melting the constituents U, Rh, Co (3N purity) and Ge (5N purity) on a water-cooled copper crucible under a high-purity argon atmosphere. A small excess of uranium (1-2 %) was used. Samples with different values of $x$ were obtained at different stages of the research. First samples with $x$ = 0, 0.2, 0.4, 0.6 were prepared, next samples with $x$ = 0.8, 0.9 and 1.0 and finally samples with $x$ = 0.93, 0.95 and 0.98. The weight loss after arc melting was always less than ~0.03 %. The as-cast buttons were wrapped in tantalum foil and annealed in evacuated ($p < 10^{-6}$ mbar) quartz tubes for 10 days at 875 ºC. The chemical composition of the annealed samples was checked by Electron Probe Micro Analysis (EPMA) for several values of $x$. The samples consisted for ~98 % of the 1:1:1 phase (*i.e.* the matrix), with the proper Rh/Co ratio. Small amounts (2 %) of uranium-rich impurity phases did form at the grain boundaries. The EPMA micrographs revealed the presence of tiny cracks in the quasi-ternary samples.

X-ray analysis of the Debye-Scherrer diffractograms for $x = 0.0$, 0.4, 0.8 and 1.0 showed the proper orthorhombic TiNiSi structure. The variation of the lattice parameters in the URh$_{1-x}$Co$_x$Ge series is anisotropic: the $a$ lattice parameter remains almost constant, while the $b$ and $c$ lattice parameter show a linear decrease with increasing $x$ [14]. The unit cell volume $\Omega$ equals 224.2 Å$^3$ for URhGe and decreases linearly at a rate of 0.162 Å$^3$/at.% Co to 208.0 Å$^3$ for UCoGe [14].

For magnetization and transport measurements bar-shaped samples (dimensions 1×1×5 mm$^3$, mass ~ 100 mg) were cut from the annealed buttons by means of spark erosion. The magnetization measurements were carried out in a squid magnetometer (Quantum Design) in the temperature range 2-300 K. The electrical resistivity was measured in the standard four-point geometry with a low-frequency ac-technique in a flow cryostat (MagLab Oxford Instruments) in the temperature range 2-300 K and in a $^3$He refrigerator (Heliox Oxford Instruments) in the temperature range 0.23-20 K. In addition, for some samples the resistivity was measured in a dilution refrigerator (Kelvinox Oxford Instruments) down to 0.02 K. Special care was taken to work with a low excitation current (< 100 µA) in order to prevent Joule heating of the samples.

## 3. Experimental results and analysis

In Fig.1 we show the temperature variation of the magnetization $M(T)$ measured in a field of 0.01 T (after cooling in a field of 1 T) and its derivative d$M(T)$/d$T$ for $x \geq 0.9$. $M(T)$ is gradually depressed and $T_\text{C}$ decreases from 8 K to 3 K for $x = 1.0$ as indicated by the minimum in d$M(T)$/d$T$. For all $x \geq 0.9$ the data below ~$0.7 \times T_\text{C}$ can be described by a phenomenological order parameter expression for ferromagnets $M(T) = M_0 (1 - (T/T_\text{C})^\alpha)^\beta$, where $\alpha$ reflects the ferromagnetic spin-wave contribution ($T < T_\text{C}$) and $\beta$ is the temperature critical exponent of the magnetization near $T_C$. Best fits are obtained with $\alpha \approx 2$, which deviates from the standard value $\alpha = 3/2$ [17], and $\beta \approx 0.3$, close to the theoretical value $\beta = 0.325$ for 3D Ising like ferromagnets [18]. The magnetization $M(H)$ measured at 2 K in a field up to 5 T is shown in Fig.2 for $0 \leq x \leq 1$. The gradual increase of $M(H)$ with increasing field is characteristic for an itinerant ferromagnet. The spontaneous magnetization $M_S$ can be obtained by fitting the data for fields exceeding ~0.5 T to the empirical function $M(H) = M_S + \Delta M(1-\exp(-\mu_0 H/B_0))$ (see Ref.19). Here $\Delta M$ determines the high-field moment $M(H=\infty) = M_S + \Delta M$ and the parameter $B_0 \sim 10$ T is a measure for the interaction strength of the fluctuating moments. The $M_S(x)$ values derived in this way are slightly larger than the field-cooled values measured in 0.01 T (see Ref.14 for $x < 0.9$ and Fig.1 for $x \geq 0.9$). Deviations from the empirical function at low field are due to demagnetizing effects ($x \leq 0.8$) or the relative large temperature (compared to $T_C$) at which the data are taken ($x \geq 0.9$).

The magnetic susceptibility, $\chi(T)$, was measured for all values of $x$ in a field $B = 1$ T in the temperature range 2 - 300 K. $\chi(T)$ is only weakly concentration dependent (see Ref.14). For $T > 50$ K $\chi(T)$ is well described by the modified Curie-Weiss law, $\chi = C/(T-\theta) + \chi_0$, where $C$ is the Curie constant. For the URh$_{1-x}$Co$_x$Ge alloys $\chi_0 \sim 10^{-8}$ m$^3$/mol, $p_{eff}$ = 1.6-1.7 $\mu_B$/U-atom and the paramagnetic Curie temperature $\theta$ attains values in the range -16 to 3 K [20]. For all samples the ratio $m_0/p_{eff}$ is much smaller than 1 which confirms the itinerant nature of the 5$f$ states in URh$_{1-x}$Co$_x$Ge [21].

The temperature dependence of the electrical resistivity $\rho(T)$ of the URh$_{1-x}$Co$_x$Ge alloys is presented in Fig.3. Note that the vertical scale has arbitrary units and the curves are shifted for clarity. The overall resistivity behaviour, namely a weak increase upon lowering the temperature and a broad maximum near 100 K, is typical for Kondo-lattice compounds (see the left panel of Fig.3). For $x \geq 0.8$ coherence effects appear before the transition to the ferromagnetic state sets in. The resistivity at low temperatures ($T < 25$ K) is shown in the right panel of Fig.3. The magnetic phase transition at $T_C$ appears as a kink in $\rho(T)$, which broadens with increasing Co concentration. The Curie temperatures (indicated by the arrows in Fig.3), were determined by the location of the maximum in d$\rho$/d$T$, and are in good agreement with values obtained by the magnetization measurements. For all samples, the resistivity drops steadily below $T_C$ where it follows the relation $\rho = \rho_0 + AT^n$, with exponent $n = 2.0 \pm 0.1$, indicated by the solid lines in the right panel of Fig.3. The coefficient $A$ attains large values and is attributed to scattering at spin waves (see next section). The residual resistivity $\rho_0$ amounts to ~80 μΩcm for $x = 0.0$. It increases up to ~300 μΩcm in the range $x = 0.4$-$0.6$ and then decreases again to ~80 μΩcm for $x = 1.0$. The large $\rho_0$ values are partly attributed to the presence of micro cracks that were revealed by the EPMA micrographs.

For $x = 1.0$ ($RRR \approx 10$) superconductivity is observed below 0.5 K, with the midpoint of the transition at $T_{sc} = 0.46$ K [2]. $T_{sc}$ is reduced to 0.41 K for URh$_{0.02}$Co$_{0.98}$Ge ($RRR \approx 6$) and no sign of superconductivity is detected in the resistivity data down to 0.05 K for $x = 0.95$.

## 4. Discussion

In Fig.4a we have traced the variation of the Curie temperature across the URh$_{1-x}$Co$_x$Ge series as determined by the magnetization and resistivity measurements. Clearly, $T_C(x)$ shows a smooth variation, with a broad maximum near $x = 0.6$ and a gradual depression towards $x = 1.0$. The variation $M_S(x)$, plotted in Fig.4b, shows a broad maximum near $x = 0.2$, but otherwise roughly follows $T_C(x)$. For $x > 0.6$, $M_S(x)$ steadily decreases as does $T_C(x)$. The overall change of the magnetic properties is also reflected in the coercive field $B_c$ which follows a similar concentration dependence. $B_c$ determined from magnetization loops measured at 2 K increases from 0.025 T for $x = 0.0$ to 0.05 T for $x = 0.4$ and then gradually decreases to small values of 0.004 T for $x = 0.9$ and 0.0003 T for $x = 1.0$ [20]. For all values of $x$ magnetization isotherms have been measured over a wide temperature range, and proper Arrott plots, i.e. plots of $M^2$ versus $H/M$, confirm itinerant ferromagnetism (see for Arrott plots of $x = 0$, 0.6 and 1.0 Ref.12, 14 and 2, respectively).

Magnetization measurements on single-crystalline samples show URhGe [12] and UCoGe [22] are uniaxial ferromagnets with the ordered moment pointing along the $c$ axis. The size of the ordered moments is 0.35-0.4 μ$_B$/U-atom [1,12] and 0.07 μ$_B$/U-atom, respectively, which is approximately twice as large as $m_0$ of the polycrystalline samples in agreement with an uniaxial anisotropy. The smooth variation of $T_C$ and $M_S$ (see Fig.4) indicates the magnetic structure remains Ising-like over the whole concentration range. The uniaxial magnetic anisotropy in UCoGe is stronger than in URhGe, where for a large magnetic field of ~12 T applied along the *b* axis the ordered moment rotates from the *c* axis towards the *b* axis [23]. Remarkably, a field-induced superconducting state is triggered by this spin-reorientation process [23].

In Fig.4b we have plotted the coefficient $A$ of the $T^2$-term in the resistivity as a function of $x$. The $A$ value spans a large range from ~1 (near $x$ = 0.4-0.6) to ~7 $\mu\Omega$cm/K$^2$ and shows a strong enhancement towards $x$ = 0 and 1. Following the Kadowaki-Woods ratio [24], we calculate a Fermi-liquid value $A_{FL}$ of 0.29 and 0.03 $\mu\Omega$cm/K$^2$, using values for the linear coefficient in the specific heat $\gamma$ of 0.170 J/moleK$^2$ and 0.057 J/moleK$^2$, for URhGe [12] and UCoGe [2], respectively. The experimental values reported in Fig.4b largely exceed $A_{FL}$, notably with a factor ~20 and ~200 for $x$ = 0 and 1, respectively. This strongly suggests the resistivity is dominated by the scattering at ferromagnetic spin waves. It also shows the abundance of low-energy magnetic excitations on approaching the ferromagnetic quantum critical point near UCoGe, in line with the scenario of magnetically mediated superconductivity [4]. It will be highly interesting to investigate in detail the anisotropy of the magnetic fluctuations in the orthorhombic structure by magnetotransport experiments on single-crystalline samples.

In summary, magnetization and transport measurements show a smooth evolution of itinerant ferromagnetic order in the pseudo-ternary series U(Rh,Co)Ge. The Curie temperature attains a maximum value of 20 K for 60 at.% Co. Electrical resistivity data show the typical Kondo-lattice behaviour. In the ferromagnetic state $\rho \sim AT^2$, with large values for the coefficient $A$, characteristic for scattering at ferromagnetic spin waves. Interestingly, the coefficient $A$ is strongly enhanced for the superconductors URhGe and UCoGe, as is expected for magnetically mediated superconductivity near a ferromagnetic quantum critical point.

Acknowledgement: The authors are grateful to S. Sakarya and N.H. van Dijk for their help during the initial stage of this work. This work was part of the research programme of EC Cost Action P16 "Emergent Behaviour in Correlated Matter".

**Figure captions**

Fig.1 (a) Temperature variation of the dc magnetization measured in a field $B$ = 0.01 T of URh$_{1-x}$Co$_x$Ge alloys for 0.9 ≤ x ≤ 1.0. Ferromagnetic order is observed for UCoGe with $T_C$ = 3 K. The solid lines represent fits to a phenomenological order parameter expression for ferromagnets (see text).
(b) Temperature derivative of the magnetization.

Fig.2 Field dependence of the magnetization of URh$_{1-x}$Co$_x$Ge alloys measured in fields up to 5 T at $T$ = 2 K. The solid lines represent fits to an empirical function $M(H)$ = $M_S+\Delta M(1-\exp(-\mu_0H/B_0))$ (see text). Co concentrations are (from top to bottom) $x$ = 0.2, 0.4, 0.6, 0.0, 0.8, 0.9, 0.93, 0.95, 0.98 and 1.0.

Fig.3 Temperature dependence of the electrical resistivity $\rho$ in arbitrary units of URh$_{1-x}$Co$_x$Ge alloys for 0 ≤ $x$ ≤ 1 as indicated. *Left panel*: 2 K ≤ $T$ ≤ 300 K. *Right panel*: 0.25 K ≤ $T$ ≤ 25 K. The Curie temperatures are indicated by arrows. The solid lines represent fits to $\rho = \rho_0 + AT^n$.

Fig.4 (a) The Curie temperature of URh$_{1-x}$Co$_x$Ge alloys as a function of Co concentration determined from $M(T)$ and $\rho(T)$ as indicated. The line serves to guide the eye.
(b) *Left axis:* The spontaneous moment $M_S$ of URh$_{1-x}$Co$_x$Ge as a function of Co concentration. *Right axis:* The coefficient $A$ of the $T^2$ term in the resistivity as a function of Co concentration.

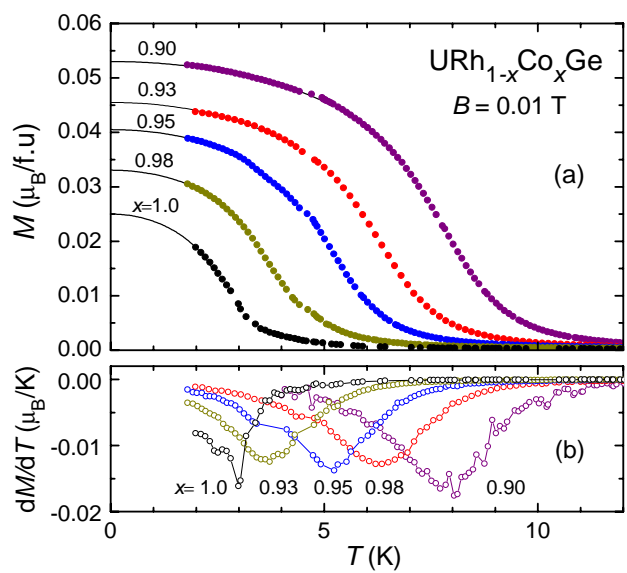

Figure 1

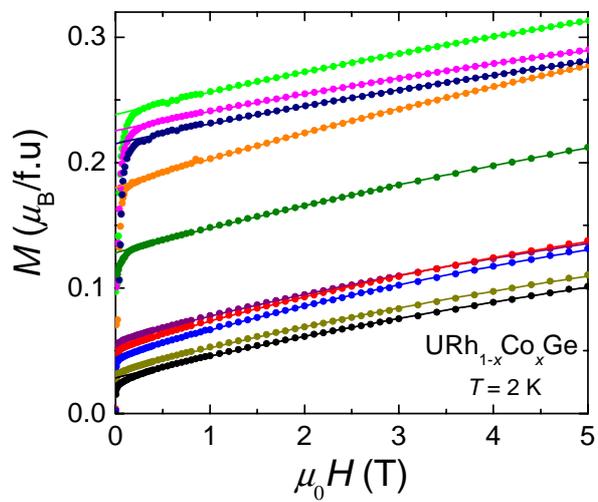

Figure 2

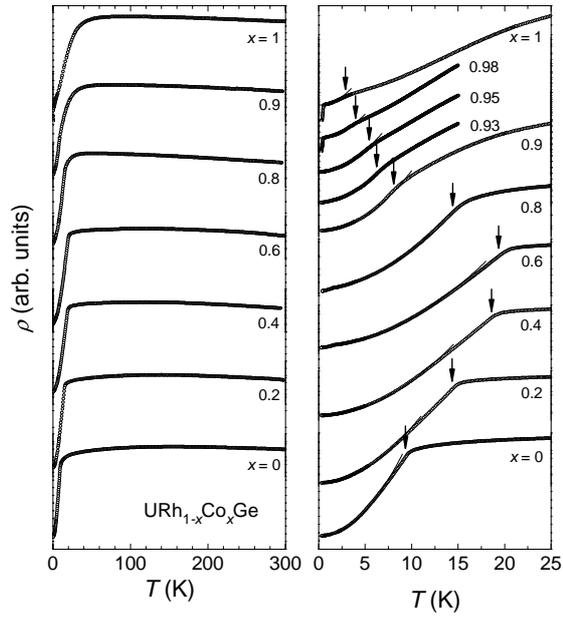

Figure 3

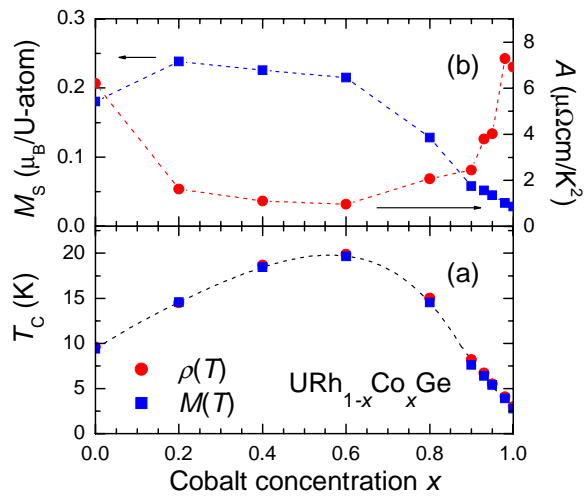

Figure 4